\documentclass[twocolumn,preprintnumbers,amsmath,amssymb]{revtex4}
\pdfoutput=1
\usepackage{graphicx}
\usepackage{dcolumn}
\usepackage{bm}

\begin{document}

\preprint{}

\title{Directional Dark Matter Detection Beyond the Neutrino Bound}

\author{Philipp Grothaus}
 \email{philipp.grothaus@kcl.ac.uk}
\affiliation{
Department of Physics, Kings College London
}
\author{Malcolm Fairbairn}
\email{malcolm.fairbairn@kcl.ac.uk}
\affiliation{
Department of Physics, Kings College London
}
\author{Jocelyn Monroe}
\email{jocelyn.monroe@rhul.ac.uk}
\affiliation{
Department of Physics, Royal Holloway University of London
}

\date{\today}

\begin{abstract}
Coherent scattering of solar, atmospheric and diffuse supernovae neutrinos creates an irreducible background for direct dark matter experiments with sensitivities to WIMP-nucleon spin-independent scattering cross sections of $10^{-46}$-$10^{-48}$ cm$^2$, depending on the WIMP mass.  Even if one could eliminate all other backgrounds, this ``neutrino floor'' will limit future experiments with projected sensitivities to cross sections as small as $10^{-48}$ cm$^2$. Direction-sensitive detectors have the potential to study dark matter beyond the neutrino bound by fitting event distributions in multiple dimensions: recoil kinetic energy, recoil track angle with respect to the sun, and event time.  This work quantitatively explores the impact of direction sensitivity on the neutrino bound in dark matter direct detection.
\end{abstract}

\maketitle

\section{\label{sec:introduction}Introduction}

Dark matter comprises approximately 25\% of the energy density of the Universe~\cite{spergel2003,Ade2013}, yet its particle properties are unknown. There are a large class of dark matter candidates with masses and interaction energy scales expected to appear just beyond the electroweak scale~\cite{snowmass2013}.  Such models are interesting since they can lead naturally to an abundance of dark matter in agreement with cosmology.  They also suggest cross sections for scattering off of nuclei which are within reach of current and near-future direct detection experiments. 

The precise predictions for scattering cross sections with nuclei in these models vary a great deal and even within a given model the scattering strongly depends upon the values of the input parameters, such that the range for the WIMP-nucleon interaction cross section $\sigma_p$ spans many orders of magnitudes.  Very small WIMP-nucleon cross sections can arise in models in which the dark matter candidate is a mixed state such that small mixing angles may suppress the interaction, or when the self annihilation cross section of dark matter in the Early Universe (which determines relic density today) is enhanced via kinematics rather than couplings.  Another possibility is that different contributions to the interaction of the WIMP with nucleons cancel each other for specific choices of input parameters. 

These possibilities require direct dark matter searches that are sensitive to very small WIMP-nucleon cross sections where, as we will see, coherent neutrino-nucleus scattering will become a problem~\cite{drukier1986,Cabrera1984,Monroe2007,Strigari2009}. 

Direct detection experiments search for dark matter particles using the coherent elastic scattering process.  Neutrinos also interact coherently with atomic nuclei, causing the nucleus to recoil with energies up to tens of keV. Such recoils would be indistinguishable from dark matter interactions individually. The scale of the ambient neutrino flux in this energy range is $10^6$ cm$^{-2}$s$^{-1}$, and the coherent neutrino-nucleus cross section is of order 10$^{-39}$ cm$^2$.  Background interactions due to these neutrinos represent a lower ``neutrino bound'' on the achievable sensitivity of dark matter direct detection experiments~\cite{Gutlein2010,Billard2013}.

Discovering dark matter with a cross section close to or below the neutrino limit will be difficult.  Equivalently, if we discover dark matter relatively soon with cross sections far above the neutrino limit we will still want to build larger detectors to study the dark matter in more detail, in which case the precision of such measurements will be limited by background neutrinos. 

In this paper, we estimate the impact of backgrounds in dark matter detectors caused by coherent neutrino-nucleus elastic scattering of ambient solar, atmospheric and supernovae neutrinos, taking into account recoil energy, direction and time modulation sensitivity.  We calculate probability distribution functions for the dark matter signal and the neutrino background in the dimensions of recoil energy, recoil direction, and event time.  We find that direction sensitivity adds approximately an order of magnitude sensitivity beyond nondirectional searches for light dark matter, and depending on the target species and energy threshold, this sensitivity can leap far beyond the solar neutrino bound.

\section{\label{sec:detectors}Dark Matter Detection}

Direct detection experiments seek to look for the signal of dark matter particles via their elastic scattering interactions with detector nuclei~\cite{gaitskell2004}.

Recent experiments have limited the magnitude of the scattering cross section to be less than approximately $10^{-45}$ cm$^2$ ~\cite{xenon100,LUX}.  This corresponds roughly to one event per 100 kilograms of detector fiducial mass per day of detector live time.   The next generation of ton-scale plus experiments are expected to increase this sensitivity by two or three orders of magnitude.

Directional detection experiments measure both the energy and track direction of the recoil nuclei.  To measure the direction of such low-energy tracks, gas targets are used at pressures of 0.05-0.1 atmospheres, with high-density readout of charge and optical signals~\cite{Ahlen2009}. Current directional detectors are at the research and development stage, and in small prototypes have demonstrated energy thresholds of a few keV, and at higher thresholds (50-100 keV) have demonstrated angular resolution of 30-55 degrees~\cite{Ahlen2009, Ahlen:2010ub, Santos:2011xk}.  For the studies here we use the event angle $\theta_{\rm sun}$, which  is defined as the angle between the recoiling nucleus track and the Earth-Sun direction.

Earlier work on directional dark matter detection has shown how dark matter properties can be constrained~\cite{Billard2010}, how exclusion limits with directional detectors may be set~\cite{Billard2010b}, how well a dark matter signal may be distinguished from an isotropic background~\cite{Morgan2004}, or how the dark matter velocity distribution may be tested~\cite{Lee2012}. We will investigate here the implications of directional dark matter detectors for dark matter searches in the presence of neutrino background.

\subsection{\label{sec:signal_xsec} Dark Matter Scattering Cross-Section }
The current approach in direct dark matter detection to test the wide range of theoretical models for dark matter is to measure the event rate of dark matter particles scattering off of target nuclei~\cite{Goodman1984}.

The zero momentum WIMP-nucleus cross section is given by
\begin{equation}
 \sigma_0=\frac{4\mu_T^2}{\pi}\left( Z f_p + (A-Z) f_n \right)^2~,
\end{equation}
where $f_p$ and $f_n$ are the couplings of the dark matter particle to the proton and neutron, respectively, $\mu_T$ is the dark matter-nucleus reduced mass, $A$ the atomic number and $Z$ the number of protons of the target nucleus. We assume $f_p$ and $f_n$ to be approximately equal such that the estimation $\sigma_0 \propto A^2 f_p^2 \mu_T^2$ holds. Then, we can cast $\sigma_0$ into the WIMP-proton cross section $\sigma_p$ via $\sigma_0 = \sigma_p \left( \mu_T / \mu_p \right)^2 A^2$ and use the event rate to constrain $\sigma_p$.

\subsection{\label{sec:phase_dist} Dark Matter Velocity Distribution}

We assume a local density for the dark matter of 0.3 GeV cm$^{-3}$ which is in agreement with current astrophysical values~\cite{fairbairn2013}.

We assume that the WIMPs have a Maxwellian distribution $f(\vec{v})$ with a cut off at the halo escape velocity $v_{\rm esc} = 544$~km/s. It is well known that limits for light dark matter depend strongly on  astrophysical uncertainties, but considering these is not part of this paper. If $|\vec{v}|<v_{\rm esc}$, the distribution in the halo rest frame is
\begin{equation}{\label{eq:vel_dis}}
f(\vec{v})_{halo} = \frac{1}{N_{\rm esc}}\left(\frac{3}{2\pi\sigma^2_v}\right)^{3/2}\exp\left[-\frac{3\left(\vec{v}\right)^2}{2\sigma^2_v}\right] ~,
\end{equation}
with $N_{esc}=\rm{erf}(z)-2z~\rm{exp}(-z^2)/\sqrt{\pi}$ accounting for the truncation. We have $z=v_{\rm esc}/\bar{v}$ and $\bar{v}=220$~km/s as the most probable WIMP velocity, which is related to the width of the distribution via $\sigma_v=\sqrt{3/2}\bar{v}$. For $|\vec{v}|>v_{\rm esc}$ we assume that $f(\vec{v})$ vanishes.

In the lab frame, we need to take into account the Earth's overall velocity vector which has contributions from the Sun's movement around the Galactic center, the peculiar movement of the Sun relative to the local standard of rest and the Earth's velocity vector relative to the sun which changes throughout the year. The velocity distribution is therefore time dependent. A detailed description of the Earth's overall velocity vector that has been used in this work can be found in~\cite{McCabe2013,Lee2013}. We integrate this time dependence over the exposure time from $t_0$ to $t_1$ to account for the annual modulation in the event rate.

\subsection{\label{sec:signal_dist} Dark Matter Signal Distribution}

The stronger the scattering cross section, the larger the event rate in an experiment. The differential rate is 
\begin{equation}
\frac{dR_{\rm DM}}{dE_r} = M_{\rm det} ~\frac{\rho_0 \sigma_0}{2m_{\rm DM}\mu^2_T}F^2(E_r) \int_{t_0}^{t_1} \int_{v_{\rm min}}^{\infty}\frac{f(\vec{v},t)}{v}d^3v~dt~.
\end{equation}
Here, $M_{\rm det}$ is the detector mass, $\rho_0$ the local dark matter density, $m_{\rm DM}$ the dark matter mass,  $f(\vec{v},t)$ the dark matter velocity distribution in Earth's frame of reference and $F(Q)$ the form factor which describes the distribution of weak hypercharge within the nucleus. The form factor depends on the momentum transfer squared, $Q^2=2 m_T E_{\rm r}$. In this work we use Helm form factors, see e.g.~\cite{lewin1995} or~\cite{horowitz1981}. To obtain the differential rate, an integral over this velocity distribution must be performed from a minimum velocity $v_{\rm min}=\sqrt{2 E_{\rm min}/m_{\rm DM}}$ that depends on the recoil energy through the minimal WIMP energy $E_{\rm min}=E_{\rm r} (m_{\rm DM}+m_T)^2/(4 m_{\rm DM} m_T)$ necessary to obtain such a recoil energy $E_{\rm r}$.

The number of dark matter events is calculated as an integral over the differential rate and the energy-dependent detection efficiency $\epsilon(E_{\rm r})$ of an experiment:
\begin{equation}{\label{eq:rate_dm}}
 s=\int_{E_{\rm thr}}^{E_{\rm up}} \epsilon(E_{\rm r}) \frac{dR_{\rm DM}}{dE_{\rm r}} dE_{\rm r} ~.
\end{equation}
If a dark matter particle with kinetic energy $E_{\rm DM}$ scatters off a target nucleus with scattering angle $\theta$ with respect to its incoming direction, the resulting recoil energy is
\begin{equation}{\label{eq:dm_Erec}}
 E_{\rm r} = E_{\rm DM} r (1-\cos \theta)/2 ~,
\end{equation}
with $r=4m_{\rm DM} m_T /(m_{\rm DM} + m_T)^2$. In this work we assume isotropic scattering in $\cos \theta$. The scattering angle of the recoiling nucleus with respect to the incoming dark matter velocity is then given by
\begin{equation}
 \tan \theta'= \frac{p' \sin \theta}{\sqrt{2 m_{\rm DM} E_{\rm DM}}-p' \cos \theta} ~,
\end{equation}
with $p'=\sqrt{2m_{\rm DM} E_{\rm DM} - 2 m_T E_{\rm r}}$.
\begin{figure}
\includegraphics[height=7.cm]{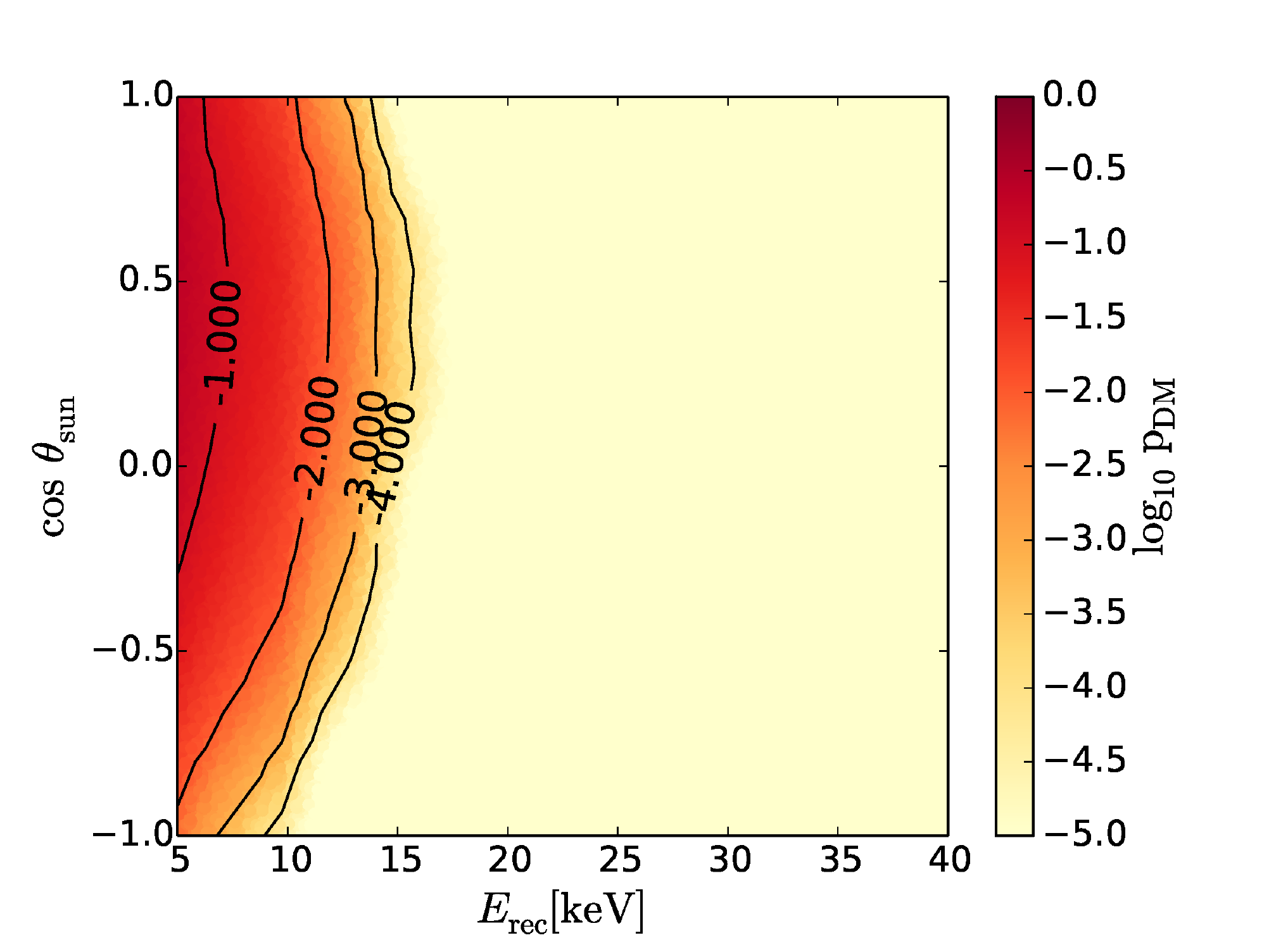}
\caption{\label{fig:dm_pdf} Two dimensional dark matter probability distribution $\rho$ of recoil energy and event angle for a 6 GeV dark matter particle in a CF$_4$ detector with 5 keV threshold in September.}
\end{figure}

Figure~\ref{fig:dm_pdf} shows the two dimensional probability distribution of event angle and recoil energy in a tetrafluoromethane, CF$_4$, detector with 5~keV energy threshold for a 6 GeV dark matter particle. Two distinct features should be noted. First, the event angles of dark matter scattering events preferably lie at large $\cos~\theta_{\rm sun}$ (small angles) because there is more solid angle (on the sphere) there. Second, the probability distribution drops to zero above the largest possible recoil energy for the given dark matter mass and escape velocity. The power of directionality is that dark matter masses that create an energy spectrum very similar to the neutrino background can easily be distinguished when the event angle is taken into account. As we will see, for light dark matter a strong gain in sensitivity compared to nondirectional detectors is therefore expected.

A third feature that is not directly visible in figure~\ref{fig:dm_pdf}, but is important nonetheless, is a variation of the peak of the dark matter probability distribution in time. The direction of the Earth's overall velocity vector will point approximately towards the radio galaxy Cygnus A~\footnote{As the direction of Cygnus A we take a right ascention of $19^{\rm h} 59^{\rm min}28.4^{\rm s}$ and a declination of $40^{\circ} 44'  1.0''$}, such that the incoming dark matter particles in the lab frame will have a preferred direction coming from Cygnus A. The relative angle between the Sun and Cygnus A changes over the year, such that the peak in the dark matter probability distribution will follow a similar pattern.

The annual modulation in the event rate of light dark matter has a maximum in June because at this time the velocity vector of the Earth and the Sun are parallel to each other~\cite{freeseannualmod}.  Both vectors approximately point into the direction of Cygnus A. In December, these two vectors are antiparallel resulting in a minimum of the event rate. The angle between the Earth-Sun direction and the Earth-Cygnus A direction, $\theta_{\rm sun-CygnA}$, is expected to be the same in June and December, because the Earth has simply moved to the other side of the Sun. However, in September the Earth is between the Sun and Cygnus A, such that $\theta_{\rm sun-CygnA}$ is at its largest value. The two objects appear on opposite directions in the sky. Analogously, in March when the Earth is behind the Sun relative to Cygnus, $\theta_{\rm sun-CygnA}$ is at its smallest value. These situations were studied to test the coordinate system of our simulations.
\begin{figure}
\includegraphics[height=7.cm]{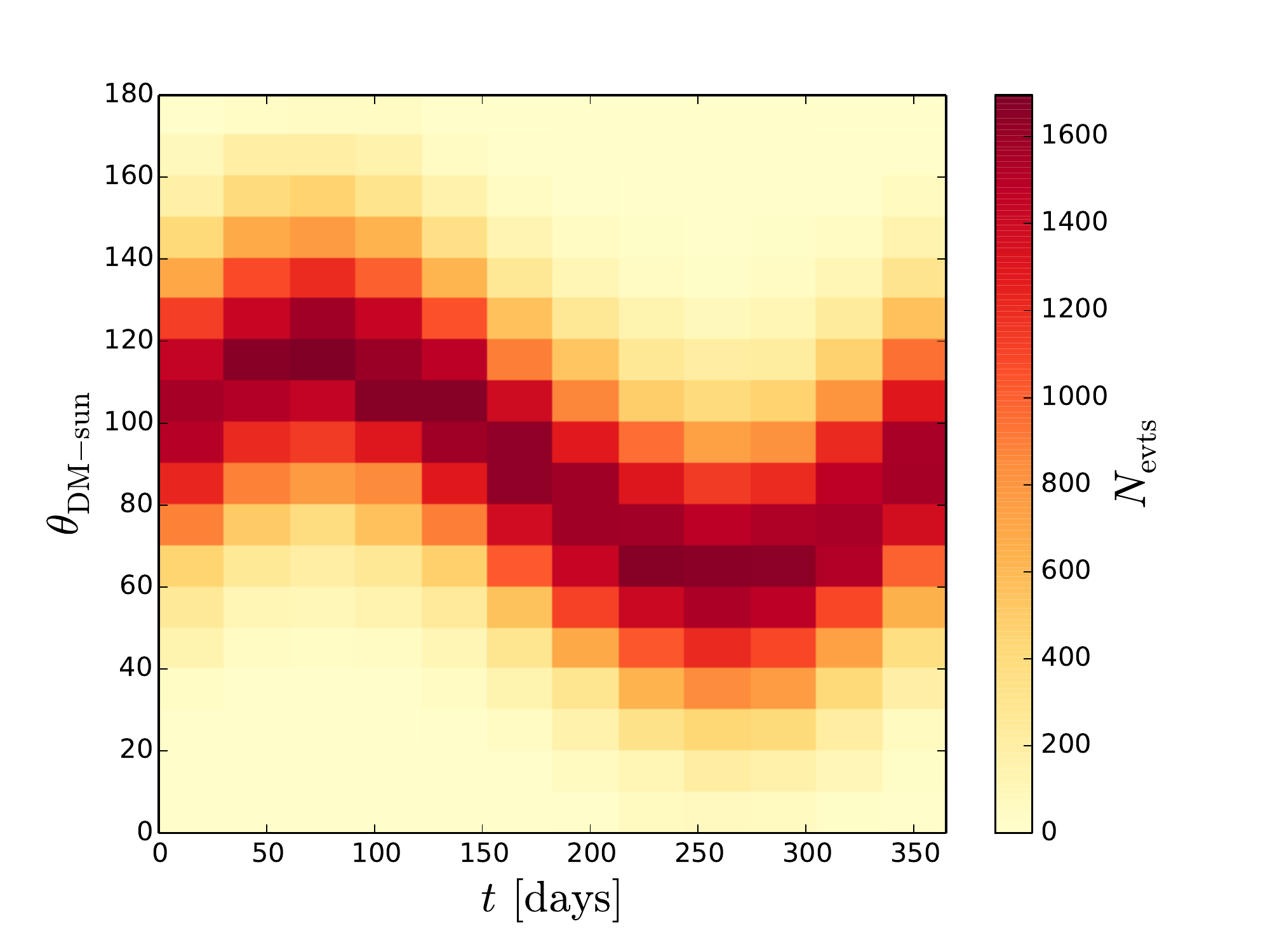}
\caption{\label{fig:dm_cygn_sun} Distribution of the angle between the incoming dark matter velocity and the Earth-Sun direction over the year for events above a 5 keV threshold in a CF$_4$ detector. For each month $1\times10^4$ dark matter events have been simulated. The maximum of the distribution follows the expected pattern as described in the text.}
\end{figure}

The time evolution of the peak in the two dimensional dark matter probability distribution arises because of this modulation in the relative angle between the incoming dark matter velocity vector and the Earth-Sun direction, $\theta_{\rm DM-sun}$. Since in September the Sun and Cygnus A appear in different directions on the sky, the velocities of the incoming WIMPs that can produce an event above a detector's fixed energy threshold therefore preferentially point along the Earth-Sun direction. In March, however, the incoming dark matter velocities will point away from the Sun, resulting in a large $\theta_{\rm DM-sun}$. When simulating light dark matter events for each month of the year and producing a histogram for $\theta_{\rm DM-sun}$, we expect the peaks of these histograms to show a modulation that follows exactly this pattern. In figure~\ref{fig:dm_cygn_sun} we color code the number of events in each angular bin. It is visible that the distribution in $\theta_{\rm DM-sun}$ follows the expected pattern with a maximum in March and a minimum in September.

Having presented the dark matter event rate as a function of energy, time and direction, we now turn to the neutrinos.

\section{\label{sec:bgnd}Neutrino Backgrounds}

Dark matter experiments are potentially sensitive to two separate types of neutrino interactions: the first is $\nu$-$e^-$ neutral current elastic scattering, where the neutrino interacts with the atomic electrons, and the second is $\nu$-$A$ neutral current coherent elastic scattering, where the neutrino interacts with the target nucleus.  The fact that the former process can lead to events in a dark matter experiment has long been realized and has led to it being suggested as a method for solar neutrino detection~\cite{bahcall1995}.  The maximum recoil electron kinetic energy from $\nu$-$e^-$ events can be as large as a few hundred keV, and the cross sections are of order $10^{-44}$ cm$^2$.  The latter process has never been observed since the maximum nuclear recoil kinetic energy is only a few tens of keV, however, the cross section is relatively large, approximately $10^{-39}$ cm$^2$.  This work focuses exclusively on coherent $\nu$-$A$ scattering.

Although coherent $\nu$-$A$ scattering has never been observed, the process is theoretically well understood. The calculated Standard Model cross section is relatively large, of order $10^{-39}$ cm$^2$~\cite{freedman1977,drukier1984}. There has been interest in using this process to make precision weak interaction measurements at the SNS~\cite{scholberg2006}, to search for supernova neutrinos~\cite{horowitz2003} and to measure neutrinos produced in the Sun~\cite{Cabrera1984}.  Even before direct dark matter detection experiments existed, this process was anticipated as a background~\cite{drukier1986}. On the other hand, one could also take the neutrino events as a signal and test neutrino physics using dark matter detectors, see e.g.~\cite{Harnik2012}.

Here we calculate the background rates caused by $\nu$-$A$ coherent scattering in target materials relevant to current dark matter searches.  We consider the recently measured solar, e.g.~\cite{Cleveland1998,Ahmed2003}, the atmospheric, e.g.~\cite{Fukuda1998,Adamson2005,Abbasi2010}, and the predicted diffuse neutrino flux from supernovae throughout the Universe and include the nuclear form factors in the coherent cross section 
calculation.  We include the direction dependence of the recoil signal, and its sidereal and annual modulation.  

\subsection{\label{sec:xsec}Neutrino Scattering Cross Sections}

The maximum recoil kinetic energy in $\nu$-$A$ coherent scattering is
\begin{equation}
E_{\rm r,max} \ = \ \frac{2 E_{\nu}^2}{m_T + 2 E_{\nu}} ~,
\end{equation}
where $E_{\nu}$ is the incident neutrino energy, and $m_T$ is the mass of the target nucleus.  The four-momentum exchange is related to the recoil energy by $Q^2$ = 2$m_T E_{\rm r}$, and the three-momentum exchange $q$ is approximately equal to $\sqrt{2 m_T E_{\rm r}}$.  For neutrino energies below 20 MeV and nuclear targets from $^{12}$C to $^{132}$Xe, the maximum recoil kinetic energy ranges between 50 and  5 keV, meaning that the maximum possible $q$ is quite small, $<$1 fm$^{-1}$.  Typical nuclear radii, $R$, are 3-5 fm, and therefore the product $q R < 1$.  In this regime, the neutrino scatters coherently off the weak charge of the entire nucleus, which is given by 
\begin{equation}
Q_W \ = \ N - (1 - 4 \sin^2 \theta_W) Z ~,
\end{equation} 
where $N$ and $Z$ are the number of target neutrons and protons respectively, and $\theta_W$ is the weak mixing angle.  Through the dependence on $Q_W$, coherence enhances the scattering cross section with respect to the single nucleon cross section by approximately a factor of $N^2$. 

The $\nu$-$A$ coherent scattering cross section is given by~\cite{Freedman1973,freedman1977,drukier1984}
\begin{equation}
\frac{d\sigma}{d(\cos \theta)} \ = \ \frac{G_F^2}{8\pi} \ Q_W^2 \ E_{\nu}^2 \ (1 + \cos \theta) \ F(Q^2)^2 ~,
\end{equation}
where $G_F$ is the Fermi coupling constant, $Q_W$ is the weak charge of the target nucleus, $E_{\nu}$ is the projectile neutrino energy, $\cos \theta$ is the scattering angle in the lab frame of the outgoing neutrino direction with respect to the incoming neutrino direction, and $F(Q^2)$ is again the nuclear form factor. The suppression of the cross section by the nuclear form factor depends on the target material and grows with the momentum transfer in a collision.

The dependence of the cross section on scattering angle means that solar neutrino elastic scattering events will, in principle, point back to the sun.  However, the majority of dark matter detectors do not have directional sensitivity, and so we calculate event rates here as a function of recoil nucleus kinetic energy as well. The scattering angle and the recoil kinetic energy are related via 2-body kinematics and the cross section can be expressed in terms of the kinetic energy, $E_{\rm rec}$, of the recoiling nucleus as
\begin{equation}
\frac{d\sigma}{dE_{\rm r}} \ = \ \frac{G_F^2}{4\pi} \ Q_W^2 \ M^2 \ (1 - \frac{m_T E_{\rm r}}{2 E_{\nu}^2}) \ F(Q^2)^2.
\end{equation}
The theoretical uncertainty on the coherent $\nu$-$A$ scattering cross section comes from uncertainty in the form factor; for neutrino energies of 10 MeV the uncertainty is expected to be less than 10\%~\cite{horowitz2003}.

\subsection{\label{sec:flux}Neutrino Fluxes}
\begin{figure}
\includegraphics[height=7.cm]{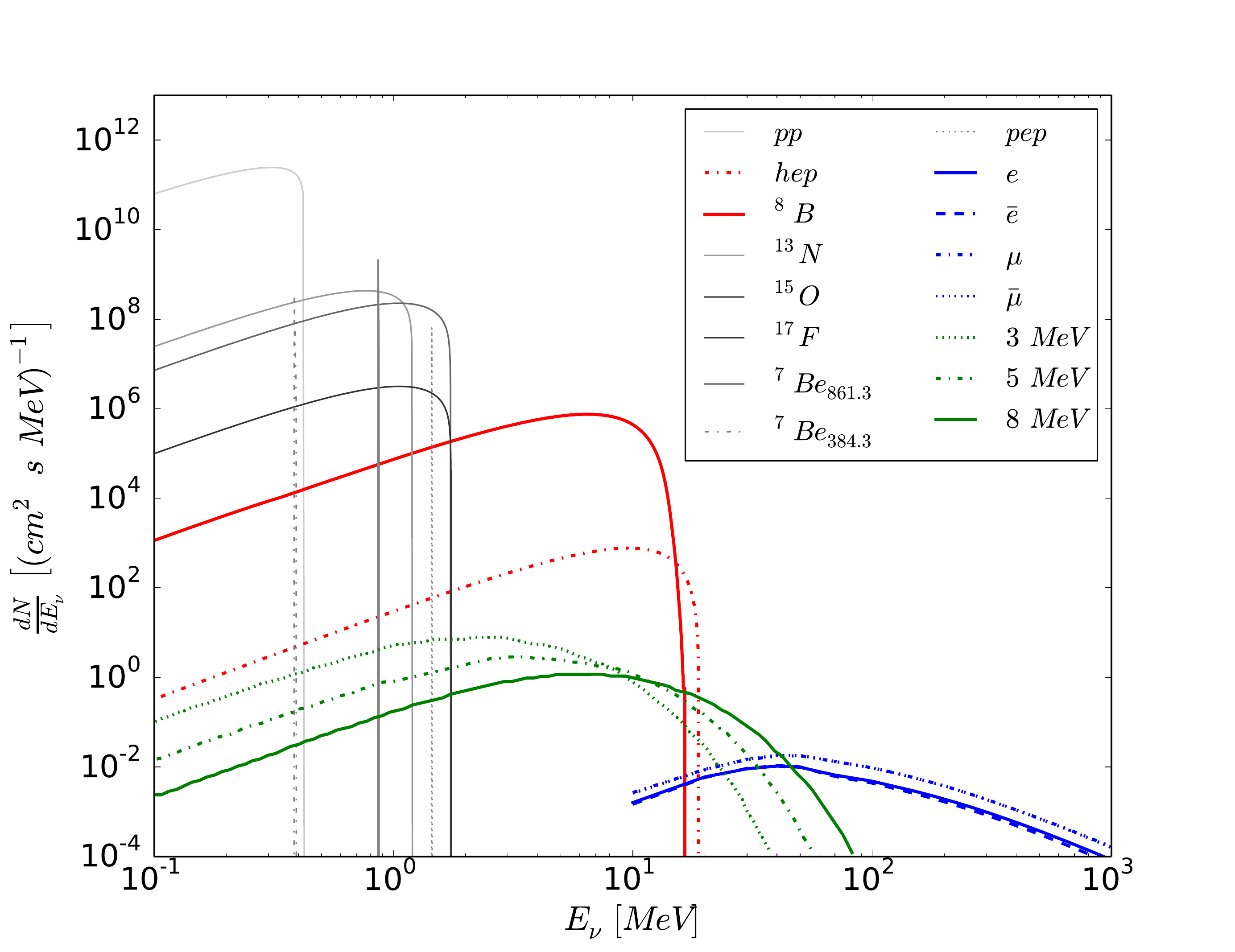}
\caption{\label{fig:flux} The neutrino fluxes considered in this work. The grey colored fluxes will not give events above thresholds considered in this paper. Important fluxes for coherent neutrino nucleon scattering originate from solar neutrinos (red), atmospheric neutrinos (blue) and diffuse supernovae neutrinos (green).}
\end{figure}

There are many sources that contribute to the large flux of ambient neutrinos and antineutrinos.  The main sources are fusion reactions in the Sun, radioactive decays in the Earth's mantle and core, decay products of cosmic ray collisions with the atmosphere, relic supernovae neutrinos and neutrinos from fission processes at nuclear reactors. We show the approximate energy ranges and fluxes of neutrinos in table~\ref{tab:flux_table}. For this work, we consider the fluxes of solar, atmospheric and supernovae neutrinos. In reference~\cite{Monroe2007} it has been shown that the contribution of geoneutrinos to the background of dark matter searches can be neglected. In figure~\ref{fig:flux} we show the energy dependent fluxes used here.

The largest contribution from solar neutrinos is the $^8$B neutrino flux, which is well understood. The predicted flux normalization, shown in table~\ref{tab:flux_table}, agrees with the measured flux at the 2\% level~\cite{sno2007}. The uncertainty of the measured flux that includes neutrino oscillations is only 3.5\%~\cite{sk2005}, even though the predicted flux normalization has an uncertainty of 16\%~\cite{bahcall2004}. For the predicted atmospheric neutrino flux the estimated normalization uncertainty is 10\% for neutrino energies below 100 MeV~\cite{superk2005} which agrees well with measurements by a number of experiments. We note that the normalization of the low energy component of the atmospheric neutrino flux is strongly dependent on the latitude. This is due to the geomagnetic cut off, e.g. the flux at Super-Kamiokande~\cite{honda2001} is approximately half of the flux at the SNO experiment.

The background of diffuse supernovae neutrinos is the integrated flux from all supernovae that occurred in the Universe. The neutrino energy spectrum of a single supernovae is assumed to be similar to a Fermi-Dirac spectrum with temperatures of 3~MeV for electron neutrinos, 5~MeV for electron antineutrino and 8~MeV for the other four flavors. For more details on diffuse supernovae neutrinos see for example~\cite{Beacom2010}. We assume an uncertainty of 10\% on the supernovae neutrino flux.
\begin{table}
\caption{\label{tab:flux_table}Ambient sources of neutrinos.  Fluxes are given in number per cm$^2$ per second.}
\begin{ruledtabular}
\begin{tabular}{lll}
Source & Predicted flux & Energy (MeV)\\
\hline
\cite{bahcall2004} Solar $\nu$ pp      & 5.99$\times10^{10}$ & $<$0.4  \\
\cite{bahcall2004} Solar $\nu$ CNO     & 5.46$\times10^{8}$  & $<$2  \\
\cite{bahcall2004} Solar $\nu$ $^7$Be  & 4.84$\times10^{9}$  & 03, 0.8 \\
\cite{bahcall2004} Solar $\nu$ $^8$B   & 5.69$\times10^{6}$  & $<$12 \\
\cite{bahcall2004} Solar $\nu$ h.e.p.  & 7.93$\times10^{3}$  & $<$18 \\
\hline
\cite{gaisser2002} Atmospheric $\nu$+$\overline{\nu}$   & O(1/E(GeV)$^{2.7})$)   & 0-10$^3$ \\
\hline
\cite{Beacom2010}Diffuse Supernovae & $T_{\nu} \approx8$ MeV & $0-10^2 $\\
\end{tabular}
\end{ruledtabular}
\end{table}

The calculations here use the predicted neutrino fluxes without including neutrino oscillations.  The coherent scattering process is neutrino-flavor independent to leading order, and we assume no sterile neutrino participation in oscillations, thus the oscillated and unoscillated predicted neutrino fluxes are, in practice, equivalent for our calculation.

\subsection{\label{sec:signal} Neutrino Signal Distribution}

The neutrino event rate is, similarly to the dark matter event rate, given by an integral over the differential recoil rate and the energy efficiency,
\begin{equation}{\label{eq:rate_nu}}
 b=\int_{E_{\rm thr}}^{E_{\rm up}} \epsilon(E_{\rm r})\frac{dR_{\nu}}{dE_{\rm r}} dE_{\rm r}~.
\end{equation}
The differential rate is
\begin{equation}{\label{eq:diffrate_nu}}
\frac{dR_{\nu}}{dE_r} = n_T ~ \int_{t_0}^{t_1} \int_{E^{\rm min}_\nu}^{\infty} \frac{dN(t)}{dE_\nu} ~ \frac{d\sigma(E_\nu, E_{\rm r})}{dE_{\rm r}}~ dE_\nu~dt ~,
\end{equation}
\begin{figure}
\includegraphics[height=7.cm]{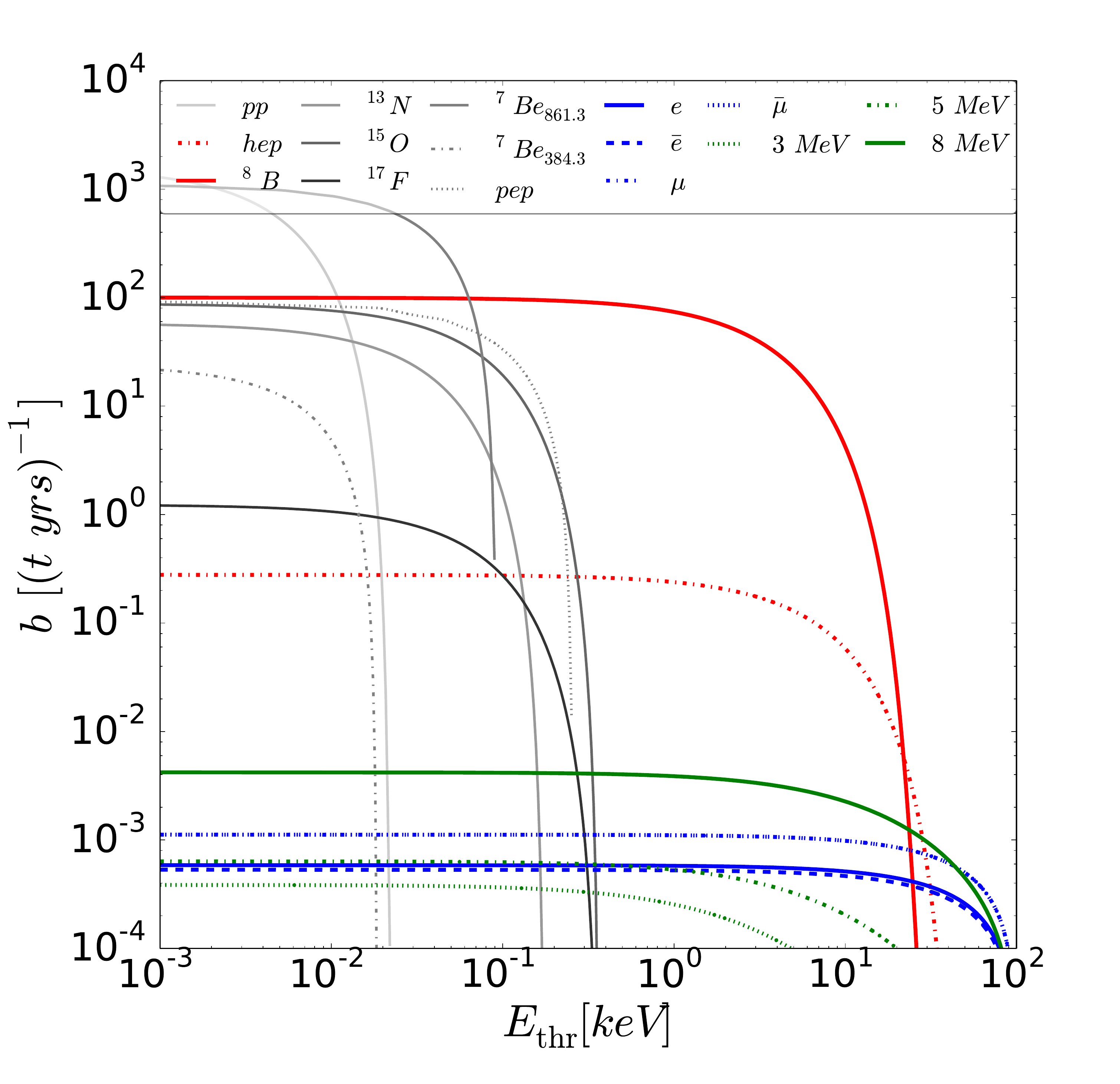}
\caption{\label{fig:cf4_rate} Neutrino event rate in a CF$_4$ detector. For this plot a perfect energy efficiency and an upper threshold of 100~keV were considered. For the rest of the paper we assume a more realistic energy efficiency function lowering the total event rate.}
\end{figure}
with $n_T$ the number of target nuclei in the detector, the flux $\frac{dN(t)}{dE_\nu}$ and the differential cross section $\frac{d\sigma(E_\nu, E_r)}{dE_r}$. The dependence on time, $t$, in the flux is due to the change in distance between the Sun and the Earth over the year. We integrate the time dependence over the exposure time of a given experiment to calculate rates. Note that the only thing that changes with time is the normalization in the solar neutrino flux, not the shape of the spectrum.  As a first approximation, we take the flux of atmospheric and supernovae neutrinos to be time independent, although there is a time variation in the atmospheric flux due to temperature changes in the Earth's atmosphere~\cite{Tilav2010}. This change in the event rate is, however, smaller than the annual modulation of the dark matter rate or the modulation of the solar neutrino rate.  As we found that both of these are not contributing significantly to the sensitivity of the simulated detectors we neglect the variation of the atmospheric neutrino flux here. 

The integral over the neutrino energy starts at the minimal neutrino energy $E_{\rm \nu}$ necessary to get a recoil event over threshold and is given by $E_{\rm \nu}^{\rm min}=\sqrt{m_TE_{\rm r}/2}$. In figure~\ref{fig:cf4_rate} we show the event rate for a CF$_4$ detector. For the threshold that we will consider in this work (5 keV), only ${}^8B$ and hep neutrinos from the Sun as well as all atmospheric and supernovae neutrinos are important.

\begin{figure}
\includegraphics[height=7.cm]{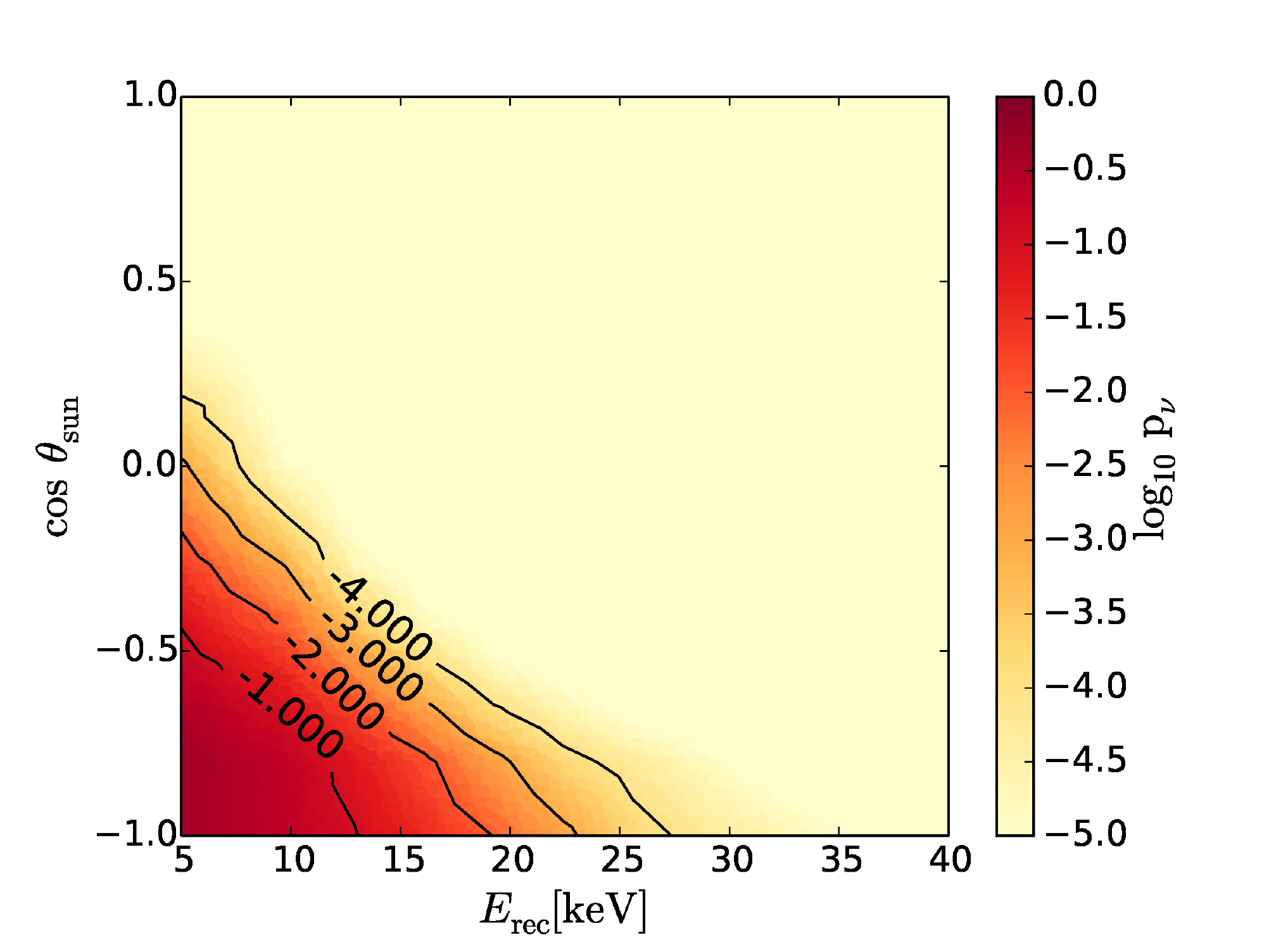}
\caption{\label{fig:neutrino_pdf} The two dimensional probability distribution $\rho$ of recoil energy and event angle of neutrinos in a CF$_4$ detector with 5~keV threshold.}
\end{figure}

The scattering angle of the nucleus with respect to the incoming neutrino direction can then be found from scattering kinematics to be%
\begin{equation}{\label{eq:scat_angle_nu}}
 \cos \theta'= \frac{E_{\nu}+m_T}{E_{\nu}}\sqrt{\frac{E_{\rm r}}{2m_T}}  ~.
\end{equation}
Figure~\ref{fig:neutrino_pdf} shows the two dimensional probability distribution of recoil energy and event angle for neutrinos in a CF$_4$ detector with a 5 keV energy threshold. The significant difference to the dark matter probability distribution is the clear peak at $\cos \theta_{\rm sun} = -1$ and small recoil energies due to the solar neutrino events. Atmospheric and supernovae neutrinos contribute as a smooth, isotropic background. For a 5 keV CF$_4$ detector we can see in figure~\ref{fig:cf4_rate} that the nonsolar neutrinos have only a small contribution such that in this example the probability distribution function falls off steeply away from the solar peak. The ratio of the solar peak to the smooth background of nonsolar neutrinos depends on the target material and the recoil energy threshold. In different detector configurations the dominance of the solar peak over the nonsolar background is not necessarily this significant.

\section{\label{sec:analysis}Dark Matter Searches in the Presence of Neutrino Backgrounds}

Having obtained detailed spectra for dark matter and neutrino events as a function of energy, direction and time, we need a statistic to test these signal and background distributions in a given experiment.  In order to do this, we perform a CLs test~\cite{Read2000} to distinguish between background and signal + background hypotheses, in which the background comes from solar, atmospheric and diffuse supernovae neutrino coherent elastic scattering.  We consider a range of targets and moderately optimistic energy thresholds, as well as energy and angular resolutions, which should be realistically achievable by the next-generation experiments.

\subsection{\label{sec:stats} Statistical Test}

The presence of backgrounds in direct searches of any kind implies that a given set of observed events  is either pure background or contains background plus signal. One way to distinguish between these two cases statistically is to perform a hypothesis test. Such a test can be carried out by looking at the ratio between the probability densities of the measured data $\vec{X}$ being either signal plus background or background only, $\widetilde{Q}=\frac{\mathcal{L}(\vec{X}, S+B)}{\mathcal{L}(\vec{X},B)}$~\cite{Read2000}. We take this as the definition of our test statistic:
\begin{equation} \label{eqn:Q}
  \widetilde{Q}=\frac{p_{b+s}(n)}{p_b(n)} \frac{\prod_{j=1}^n\frac{sS_t(t_j)+bB_t(t_j)}{s+b}~\frac{sS_{\theta,E}^{(t)}(\theta_j,E_j)+bB_{\theta,E}(\theta_j,E_j)}{s+b}}{\prod_{j=1}^n B_t(t_j)~B_{\theta,E}(\theta_j,E_j)} ~.
\end{equation}
Throughout this work, we use the notation $p(x)=dP(x)/dx$ as the probability distribution function of the variable $x$ where $P(x)$ is therefore the cumulative probability of this quantity at $x$.  In equation~\ref{eqn:Q}, $s$ is the number of expected dark matter events given by equation~\ref{eq:rate_dm}, $b$ the number of expected neutrino events given by equation~\ref{eq:rate_nu} and $n$ the total number of observed events in an experiment. The functions with capital letters $B$ or $S$ denote different normalised probability distribution functions for the neutrino and dark matter events, respectively. $p_{\lambda}(n)$ is the Poisson distribution centered at $\lambda$, where $\lambda$ is either $b$ or $b+s$ (we discuss in section~\ref{sec:uncertainties} how to exactly obtain $s$, $b$ and $n$). The variables $t_j, E_j, \theta_j$ denote the time, recoil energy and event angle of the j-th event. We define the event angle as the angle between the track of the recoiling nucleus and the Earth-Sun direction.

$B_t$ describes the annual modulation of the neutrino event rate. It has a maximum in January, when the neutrino flux and hence the neutrino event rate is largest, and a minimum in July, when the distance between the Earth and the Sun is at its maximum.  $S_t$ encodes the information of the annual modulation of the dark matter event rate and depends on the dark matter mass.  For light dark matter this function has a maximum in June and a minimum in December.

$B_{\theta,E}$ is the two dimensional probability distribution of the recoil energy and the event angle for neutrino events and $S_{\theta,E}^{(t)}$ the corresponding one for dark matter events. Visualized examples of these distributions are the figures~\ref{fig:neutrino_pdf} and~\ref{fig:dm_pdf}, respectively. The dark matter distribution carries an additional index for time because of its variation over the year as described in section~\ref{sec:signal_dist}. To include this time variation, we choose ten equally distributed days over one year and create one probability distribution function for each of these days. A given event will then interpolate linearly between the two probability distribution functions closest to the signal event time. Equation~\ref{eqn:Q} can be simplified to:
\begin{equation} \label{eq:simpleQ}
  \widetilde{Q}=e^{-s}\left(\frac{b}{s+b}\right)^n \, \prod_{j=1}^n \left(1+\frac{sS_t(t_j)}{bB_t(t_j)} \right) \left(1+\frac{sS_{\theta,E}^{(t)}(\theta_j,E_j)}{bB_{\theta,E}(\theta_j,E_j)}\right) ~.
\end{equation}
In the following we will discuss the log-ratio $Q=-2\log\widetilde{Q}$.

An advantage of this procedure is that experimental uncertainties can easily be incorporated by smearing the probability distributions. Dark matter searches have to deal with imperfect energy and angular resolution (in the case of directional experiments), as we discuss in section~\ref{sec:detassumptions}, leading to a smearing of $B_{\theta, E}$ and $S_{\theta,E}^{(t)}$. The background of nonsolar neutrinos ensures a non-zero value for $B_{\theta, E}$ for all values of $\theta$ and $E$ such that $Q$ is well behaved. See section~\ref{sec:ev_sim} for more details on how $B_{\theta, E}$ and $S_{\theta,E}^{(t)}$ are created.

For every dark matter mass and cross section we want to find out wether a fixed detector setup (target material, energy threshold, exposure, energy and angular resolution) is capable of distinguishing whether the observed events are pure background or contain a dark matter signal. To do so, Q has to be evaluated twice: First we simulate pseudoexperiments with only neutrino events and obtain a distribution $p_B(Q_B)$ for the background only hypothesis using equation~\ref{eq:simpleQ}. As, in this case, the pseudodata is more consistent with the background expectations, $p_B(Q_B)$ will peak at positive Q values.

We then repeat the exercise and simulate pseudoexperiments with dark matter and neutrino events to get a distribution $p_{SB}(Q_{SB})$. This distribution will, in contrast, peak at negative Q values.

\begin{figure}
 \includegraphics[height=7.cm]{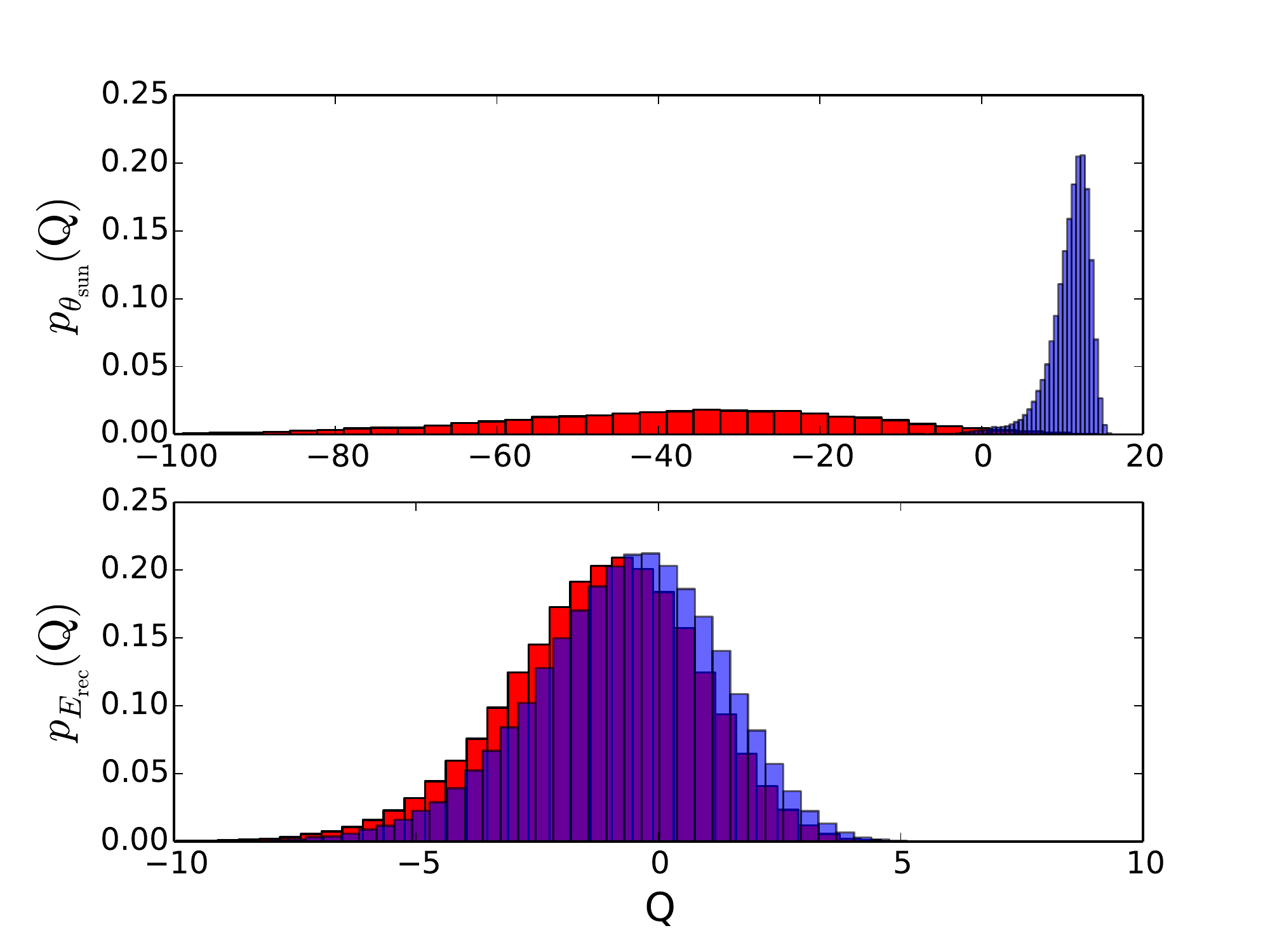}
\caption{\label{fig:Q_dis} The normalised background only distribution  $p_B(Q_B)$ (blue) and signal plus background distribution $p_{SB}(Q_{SB})$ (red) including angular information (top) and excluding angular information (bottom) for s=10 and b=500 for a 6~GeV dark matter particle in a CF$_4$ detector. The gain in sensitivity when using directionality is clearly visible in the separation of the two distribution in the upper plot.}
\end{figure}

We can then decide whether the detector setup is sensitive to a given $m_{\rm DM}$ and $\sigma_{\rm p}$, if we look at the separation between these two distributions. The clearer this separation, the easier it is for the chosen detector configuration to distinguish between these two hypotheses and the more sensitive the detector. The lower the signal rate and the more similar the signal expectations  are to the background expectations, the closer the distributions will be until they start to overlap. If the overlap becomes too large, the experiment will lose its sensitivity completely. See figure~\ref{fig:Q_dis} for a visualization.

To quantify the sensitivity of a given dark matter experiment for a specific dark matter mass and cross section we calculate the overlap of these two distributions as follows. We integrate both,
\begin{equation}
 \beta_{SB}=\int_{-\infty}^{q} p_{SB}(Q_{SB})~dQ_{SB}~,
\end{equation}
\begin{equation}
 \beta_{B}=\int_{-\infty}^{q} p_{B}(Q_{B})~dQ_{B}~,
\end{equation}
up to a $q$ value for which
\begin{equation}
1-\beta_{SB}=\beta_B\equiv \alpha ~.
\end{equation}
We take the confidence level at which the signal plus background hypothesis can be distinguished from the background-only hypothesis to be $(1-\alpha)$. In this work we are interested in separations of both hypotheses at $90\%$ confidence level, corresponding to $\alpha$ equal to 0.1,  and in 3$\sigma$ separations ($\alpha=0.00135$).

The statistical approach has uncertainties due to a finite sample size of pseudoexperiments, a finite number of events to create the two-dimensional probability distributions, as well as a finite bin width when creating the histograms of the test statistics. We estimate this numerical error to be 5\% in the overlap and add it to the error due to the systematic uncertainties.

\subsection{\label{sec:detassumptions}Detector Performance Assumptions}

In this work we will estimate future sensitivities of dark matter detectors including the irreducible neutrino nucleus coherent scattering as a background. To see how the mass of the target material influences the sensitivities, we look at tetrafluoromethane, CF$_4$, as a light and Xenon, Xe, as a heavy target material. 

Interesting directional technologies are already in existence for experiments based on CF$_4$~\cite{Ahlen2009,Monroe2011} and we show here that scaling these detectors to large masses can test cross sections beyond the neutrino background. For Xenon, on the other hand, there are at the moment no directional techniques demonstrated and our directional sensitivities are in this sense futuristic. However, we think it is still interesting to see how directional information would help if a heavy target material was used.

For the CF$_4$ detectors, we model the energy efficiency of the detectors with
\begin{equation}
 \epsilon(E_{\rm rec}) = c_1\left(1 + {\rm erf}\left[ \frac{(E_{\rm rec} - c_2)}{c_3} \right] \right) ~,
\end{equation}
and choose $c_1=0.5$, $c_2$ as the energy threshold $E_{\rm thr}$ and $c_3=15$~keV. These values for the efficiency asymptote at 50\% and are consistent with current directional searches~\cite{Ahlen2009}. The 5 keV energy threshold we assume is optimistic relative to current searches (although a number of CF$_4$ directional detectors use 5.9 keV $^{55}$Fe sources for calibration, and track images with directionality at this energy have been measured in small prototypes~\cite{Santos:2011xk}).  Large direct dark matter searches based on Xenon have been carried out in the past already~\cite{LUX,xenon100}, so we use the efficiency curve published by the LUX experiment and shift it to smaller energy thresholds, such as 2 keV. For the Xenon detectors we assume an upper energy window cutoff of 40~keV, for the CF$_4$ detectors we take 100 keV.

The energy resolution is modeled as
\begin{equation}
 \sigma_E = 0.1~\sqrt{E/{\rm keV}}~,
\end{equation}
and the angular resolution as
\begin{equation}
 \sigma_\theta = \frac{30^{\circ}}{\sqrt{E/{\rm keV}}} ~.
\end{equation}
For the angular efficiency we assume $100\%$.

\subsection{\label{sec:ev_sim} Event Simulation}

To simulate dark matter events we use the rest frame of the static, spherically symmetric dark matter halo and draw a random velocity magnitude $v$ from the velocity distribution according to equation~\ref{eq:vel_dis} and, to fix the dark matter direction, additionally two angles ($\theta$,$\varphi$) in a spherically symmetric way. We then calculate the cartesian coordinates of the dark matter velocity vector $\vec{v}$ in galactic coordinates. Drawing a random event time $t$ from a uniform distribution between $t_0$ and $t_1$ gives us the Earth's overall velocity vector in galactic coordinates from reference~\cite{McCabe2013}. After a coordinate transformation into the rest frame of the Earth, we have the incoming WIMP velocity vector. 

As we assume isotropic scattering, we draw a uniform scattering angle and obtain the recoil energy $E_{\rm r}$ of the event from equation~\ref{eq:dm_Erec}. In this way, the dark matter direction and annual modulation are both included in the event simulation as we use the full information of the Earth's velocity vector and start from a spherically symmetric halo. We calculate the event angle $\theta_{\rm sun}$ by projecting the track of the recoiling nucleus onto the Earth-Sun direction and perform the energy and angular Gaussian smearing. To take the energy efficiency into account, we only accept the corresponding fraction of events at each recoil energy and apply the energy thresholds as hard cutoffs. For each of the ten dark matter probability distributions of each dark matter mass we simulate $10^6$ events and bin the data into 30 energy and 15 angular bins.

To create the two dimensional probability distribution function for neutrinos, we perform the event rate calculation, equation~\ref{eq:rate_nu}, for each neutrino type separately to know exactly how many events of each type can be expected in a given detector configuration. 

To simulate neutrino events, we draw a random neutrino energy according to the energy dependent flux. For a solar neutrino the direction is known. When simulating the atmospheric and supernovae neutrinos we assume an isotropic incoming neutrino direction. We use the differential cross section and its dependence on the neutrino energy to create a probability distribution for a given neutrino energy $E_{\nu}$ to give an event of recoil energy energy $E_{\rm r}$. From this we draw a random $E_{\rm r}$ and obtain the scattering angle via equation~\ref{eq:scat_angle_nu}. 

These real event values are smeared according to the detector resolutions as explained in section~\ref{sec:detassumptions} and the energy efficiency and thresholds are applied. The event time is drawn uniformly for the atmospheric and supernovae neutrinos, but from a non uniform distribution for the solar neutrinos that follows the annual modulation of the event rate. For the neutrino probability distribution we simulate $1.5\times 10^6$ neutrino events and bin them into 30 energy and 15 angular bins.
\begin{figure}
\includegraphics[height=7.cm]{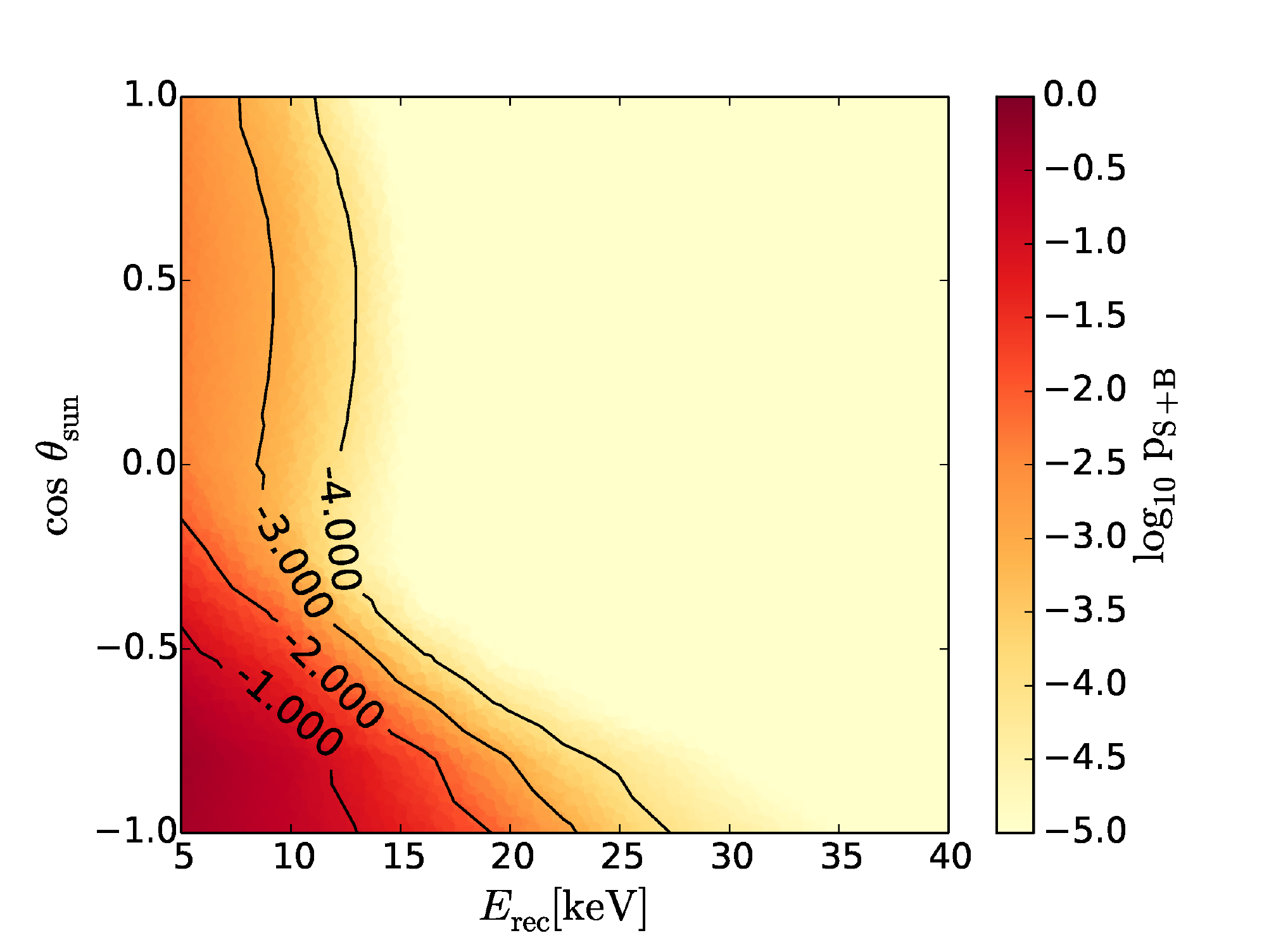}
\caption{\label{fig:combo_pdf} The combined two dimensional probability distribution $\rho$ of the recoil energy and event angle for a 6 GeV dark matter particle and neutrinos in a CF$_4$ detector. The expected signal rate is fixed to s=10 and the expected background rate to b=500.}
\end{figure}

In figure~\ref{fig:combo_pdf} we present the combined two dimensional signal plus background probability distribution of event angle and recoil energy for a 6 GeV dark matter particle in a CF$_4$ detector. As can be seen in equation~\ref{eqn:Q}, the constituent probability distribution functions of signal and background are weighted according to the expected number of background and signal events. We present the case for b=500 and s=10. Even for such low count rates, a significant excess at large $\cos \theta_{\rm sun}$ is visible, compared to figure~\ref{fig:neutrino_pdf}.

Given the differences between the energy, angle, and time distributions of dark matter signal and neutrino background events, a directional experiment could in principle fit simultaneously for the normalizations of both fluxes and include the systematic uncertainties as nuisance parameters.  This technique has been employed by a number of experiments to constrain systematic uncertainties on both background and signal distributions, for example the current most precise measurement of the solar $^8$B flux~\cite{Aharmim2009}.  However, the degree to which this approach is successful depends strongly on the number of events, the separation of the signal and background distributions (which in the energy dimension depends on the dark matter mass), and the degree of correlation of the systematic uncertainties.  Therefore we assume for this work that the neutrino flux uncertainty is externally determined, and we make a semioptimistic assumption about its magnitude in the future.

\subsection{\label{sec:uncertainties}Systematic Uncertainties}

We have seen that the neutrino background is made up of three distinct populations - solar neutrinos, diffuse neutrino background from supernova explosions and neutrinos from cosmic rays hitting the atmosphere.
The amount by which the uncertainties on these three fluxes will reduce in future years is uncertain and can only be estimated roughly.  The solar flux in our energy range of interest is dominated by $^8$B and hep neutrinos, and the flux of neutrinos due to both of these emission mechanisms is very sensitive to the iron abundance in the Sun which affects the opacity in the core~\cite{PenaGaray2008}.  Understanding the iron opacity in the Sun is challenging - currently the solar composition as observed at the surface of the Sun~\cite{Asplund2009} is not in good agreement with that deduced from helioseismology~\cite{2014ApJ787}.  There are at least two future experiments that will help reduce the uncertainty on the solar flux.  SNO+ will make good measurements of both the Boron-8 and the Beryllium-7 neutrinos both of which depend sensitively upon the iron abundance in the core, which will indirectly constrain the hep fluxes~\cite{Maneira2013}. If approved, Hyper-Kamiokande should be able to detect several hundred $^8$B neutrinos per day collecting such high statistics that it will look for time variation in the $^8$B flux~\cite{Abe2011}.  With these observations, it seems not impossible that the uncertainties in the solar flux may drop by a factor of several in a few decades, if not orders of magnitude. See the recent work~\cite{Billard2014} for how data from future dark matter detectors could help to test solar models.

The diffuse supernova background (DSNB) neutrinos are more complicated since while we are able to increase our understanding of the historic stellar evolution history using astronomical observations, we are not so certain of the spectrum of neutrinos emitted from a single supernova.  It seems that there are good prospects for Super-Kamiokande to improve its sensitivity, in particular by using dissolved gadolinium to improve neutron tagging, which could significantly enhance its sensitivity to the DNSB neutrinos ultimately leading to a discovery in a decade or so of running which could constrain the magnitude of the spectrum between 10-20 MeV~\cite{Beacom2013}.  Again, Hyper-Kamiokande would do much better, allowing one to measure the spectrum in great detail.  It is not unreasonable to suggest a large drop in the uncertainties in this flux.

The same enhancement of Super-Kamiokande could detect the atmospheric neutrino flux.  This flux is perhaps more elusive than the other two since factors which may affect it include the primary cosmic ray flux (although this will be constrained by AMS02), the geomagnetic field, the solar wind and nuclear propagation models.  The theoretical uncertainties in the development of the shower have been studied by observing the interaction between protons and thin targets of $O_2$ and $N_2$ at the HARP experiment in CERN as well as observing the muon flux in the atmosphere using balloon experiments.  Models like DPMJET-III and JAM then aim to constrain the resulting neutrino flux~\cite{Honda2011}.  Because of this, the uncertainties are likely to fall more slowly, although Hyper-Kamiokande and an upgraded Super-Kamiokande will probably be able to detect the flux therefore constraining it more tightly.

These considerations lead us to the semioptimistic approach to take half the current flux uncertainties as a basis for the simulated detectors in our analysis. This translates into an uncertainty of 8\% for the solar neutrino fluxes (note that only $^8$B and hep neutrinos can give events above threshold in this work), and 10\% for the atmospheric and supernovae neutrino fluxes.

To include these neutrino flux uncertainties we first obtain a central value result. This means that we assume the incoming fluxes to have their nominal measured values resulting in a background rate $b_0$. The number of observed events $n$ in a pseudo experiment is drawn from a Poisson distribution centered at a value $\lambda$ which is either equal to $b_0$ for the background only or $b_0+s$ for the signal plus background simulation. For each pseudo experiment we simulate these $n$ events as we discussed in section~\ref{sec:ev_sim}. 

To account for the unknown real flux value when performing the experiment we vary the expectation of each pseudo experiment, that is $b$ in equation~\ref{eq:simpleQ}. Hence, for each pseudo experiment we draw a random flux value for each neutrino flux type from a Gaussian with 1$\sigma$ corresponding to the uncertainties. This results in a different expected background rate $b$ for each pseudoexperiment via equation~\ref{eq:diffrate_nu} and widens the $Q$ distributions. We then repeat the procedure shifting $b_0$ up and down by one sigma to obtain a 1 sigma band for the estimated exclusion limits.

\section{Results}

\subsection{\label{sec:results} Estimation of Detector Sensitivities}
 
In order to see directly the gain in sensitivity when directional information is used, we evaluate the sensitivity that we obtain from our statistical approach for both cases, excluding (red bands) and including directional information (green bands). To compare the results to the WIMP discovery limit that was presented in~\cite{Billard2013}, we show this limit as a light grey line. Note here that the limits from~\cite{Billard2013} are discovery limits at the $3\sigma$ level and based on a profile likelihood approach, whereas we perform a hypotheses test. Therefore, any direct comparison should be taken with care. A strict discovery limit exists for dark matter masses that match the energy spectrum of the neutrino background perfectly, see~\cite{Billard2013}. This is, for example, the case for a 6 GeV dark matter particle and the background of $^8$B neutrinos in a Xenon detector. We reproduce this limit and the discovery limits for heavy dark matter from~\cite{Billard2013} with very good accuracy; see also section~\ref{sec:max_sensitivity}. In the dark matter mass region around 10 GeV where a steep increase in sensitivity towards smaller cross sections is observed, however, we find slightly less constraining discovery limits, as will become clear when we discuss the Xenon detector.

In this section we will look at sensitivity limits at the 90\% C.L. and 3$\sigma$ level for experiments with different target materials and energy thresholds. To compare the different simulations, the detector exposure is scaled such that the simulated experiment will observe 500 neutrino events, i.e. the background contribution is sizable. As an example for a dark matter detector with directionality, we estimated the sensitivity of tetrafluoromethane CF$_4$ as target material. As a light target CF$_4$ is promising to distinguish solar neutrinos from light dark matter. We set the energy thresholds in our run to 5~keV.
\begin{figure}
\includegraphics[height=7.cm]{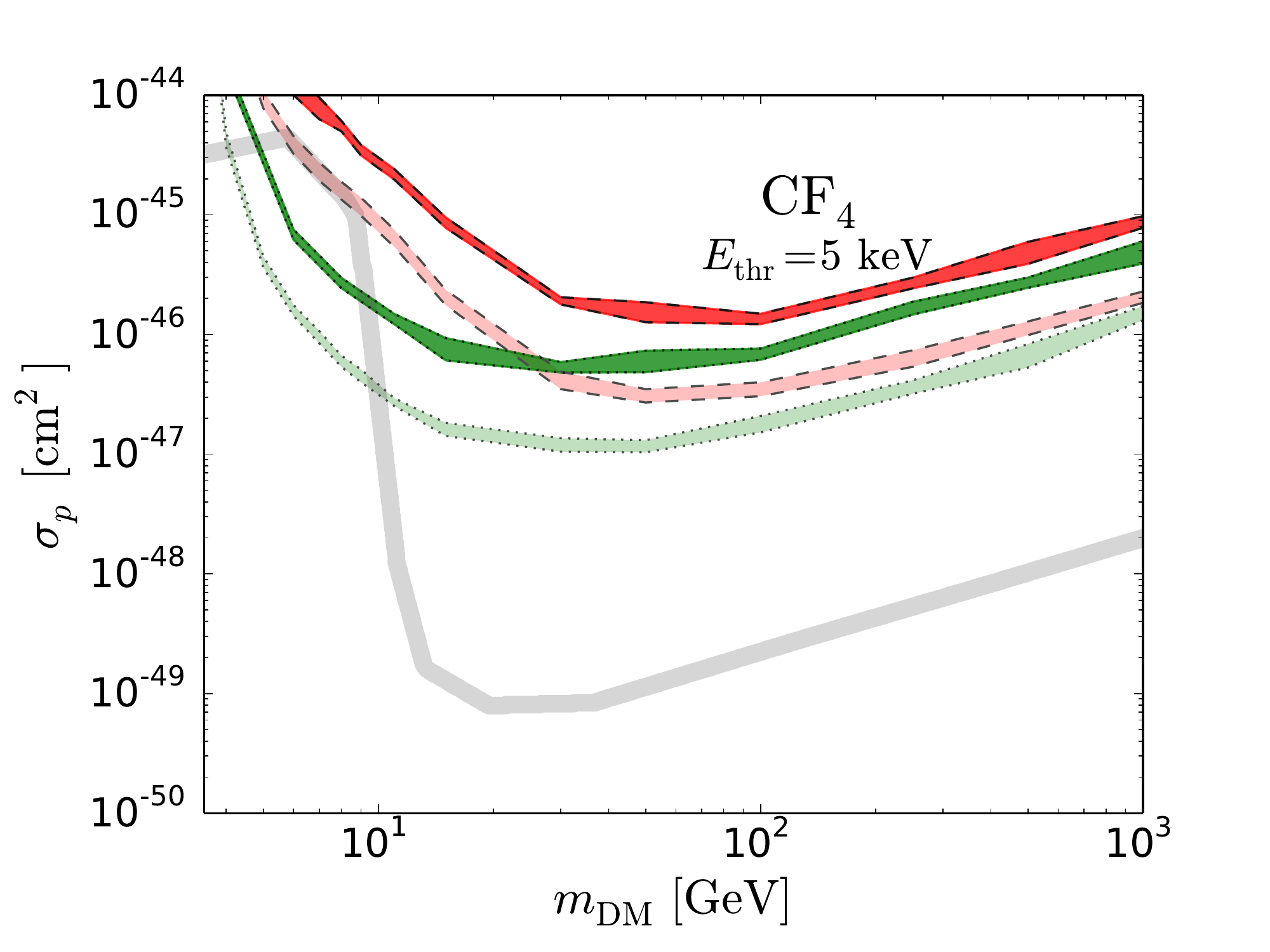}
\caption{\label{fig:cf4_run1} Estimated sensitivity limits at 3$\sigma$ level for a nondirectional (red band) and directional (green band) CF$_4$ detector with 36 t-yrs exposure and 5 keV energy threshold resulting in 500 expected neutrino events. The fainter bands indicate corresponding sensitivity limits at 90\% CL, the grey curve is the neutrino bound. }
\end{figure}

Figure~\ref{fig:cf4_run1} shows the obtained sensitivity bands for a 36.6 ton-year CF$_4$ experiment with a 5~keV energy threshold. The 500 neutrino events consist of 499.8 expected solar and 0.2 expected nonsolar neutrinos. The green and red bands represent limits that can be obtained with directional and nondirectional detectors at a 3$\sigma$ level, respectively, and the grey curve is the neutrino bound. The fainter colors show corresponding limits at 90\% C.L. The separation of the green band from the red band clearly shows the impact of directional information. A strong increase in sensitvity for directional detectors towards smaller cross sections is observed which is larger the smaller the dark matter mass. This is easily understood when considering the clear seperation of the neutrino and dark matter peak in the two dimensional probability distribution functions. The lighter the dark matter particle is, the more significant this separation. For a light dark matter event to be above threshold, the track of the recoiling nucleus has to lie closer along the incoming dark matter direction in order to produce a large enough recoil. Hence, the dark matter signal also has a strong directional character, as discussed in section~\ref{sec:detectors}. Since the event angle distribution is different for the neutrinos, directional information has a large impact. 

We find that cross sections below the solar neutrino bound can be tested at the  3$\sigma$ level when directional information is taken into account.

Towards heavier dark matter masses, we see that the sensitivity curves approach each other and directionality loses some impact. For heavy dark matter, the distinction of signal and solar background is already easy when the energy spectrum is considered on its own, because the recoil energies of solar neutrinos are much smaller compared to heavy dark matter. Besides, the dark matter events lose their directional character more and more: Light dark matter can only give recoil energies above threshold for the largest dark matter velocities in the halo, such that only those particles coming from Cygnus A can give a recoil event in the detector. The kinetic energy of heavy dark matter particles is, in contrast, also large for small dark matter velocities. Hence, the incoming direction of dark matter particles that give a signal event in the detector becomes unconstrained and more and more isotropic. A competing effect is that the track resolution for small recoil energies is worse, but improves for larger recoil energies and thus for heavier dark matter. Overall, we see that directional information is also useful for heavier dark matter. This is mainly because when heavy dark matter particles give recoil energies comparable to the recoil energies of solar neutrinos, the dark matter events can be distinguished using directional information, which would not be possible otherwise. 
\begin{figure}
\includegraphics[height=7.cm]{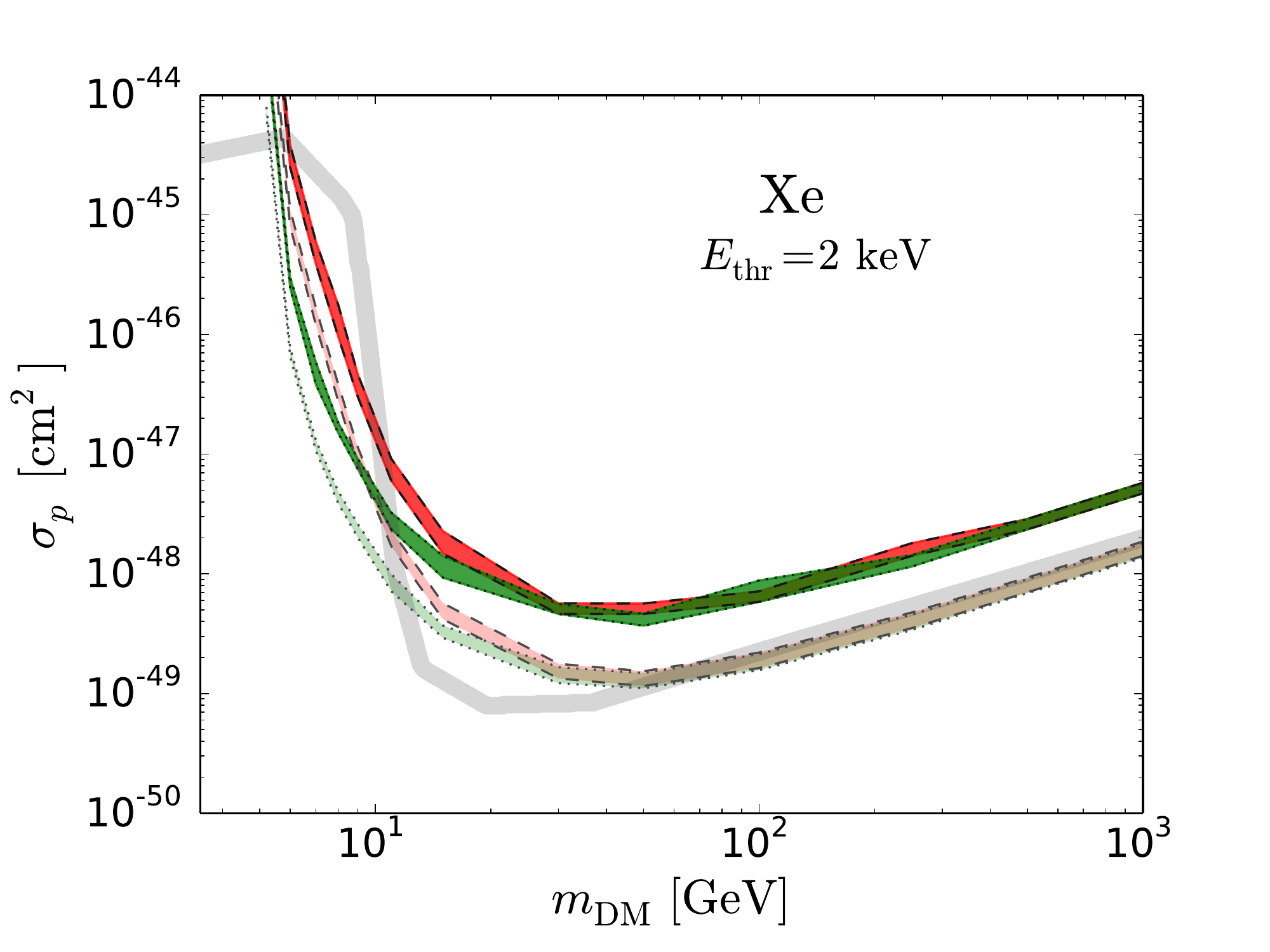}
\caption{\label{fig:Xe_run1} Estimated sensitivity limits at 3$\sigma$ level for a nondirectional (red band) and directional (green bands) Xenon detector with 367 t-yrs exposure and 2 keV energy threshold resulting in 500 expected neutrino events. The fainter bands indicate corresponding sensitivity limits at 90\% CL, the grey curve is the neutrino bound. }
\end{figure}

At the moment, the strongest constraints on the WIMP-nucleon cross section are set by experiments that use Xenon as a target material. These detectors have no directional information and no technology exists up to now that could achieve this. However, it is still interesting to ask which cross section experiments with heavy target materials would be able to probe if they could use directional information.  There has been recent interest in developing a direction-sensitive Xenon detector technology based on recombination dependence on the recoil angle relative to the detector $\vec{E}$ field~\cite{Nygren:2013nda}, so perhaps this will be a possibility for the future.

Therefore, we additionally choose Xenon as a target material and perform the same tests. Estimated sensitivity curves for a hypothetical experiment with 367.7 ton-year exposure using a 2~keV threshold can be seen in figure~\ref{fig:Xe_run1}. The 500 neutrino background events consist of 485.8 expected solar and 14.2 expected nonsolar neutrinos. 

Our statistical test finds that even without directional information cross sections below the discovery limit from~\cite{Billard2013} (grey curve in the plots) can be tested at 3$\sigma$ level. For example, an 8~GeV WIMP with a cross section of $2.3\times10^{-46} \rm{cm}^2$ would give about 470 dark matter events. We note here, that we assumed half the flux uncertainties and took a different statistical approach than reference~\cite{Billard2013}. The nondirectional 3$\sigma$ limit should hence be seen as a WIMP-discovery limit obtained from our approach rather than testing cross sections beyond the discovery limit. Again, we see that directional detectors can go beyond and probe smaller cross sections compared to nondirectional detectors. The same trend that directional and nondirectional detectors give similar sensitivities for heavy dark matter particles is visible; the limits are basically identical for the Xenon detector. 

Compared to the light target material CF$_4$ we find that the impact of directional information is less significant in this Xenon detector configuration when searching for heavy dark matter. With Xenon as a heavy target material solar neutrinos can give recoil energies only up to approximately 5~keV. Hence, the range of recoil energies for which directionality is the only indicator to distinguish the signal from the solar neutrino background is small. For the light target material CF$_4$ this range is larger: solar neutrinos can recoil up to approximately 30~keV, see figure~\ref{fig:cf4_rate}. We can therefore conclude that the larger the range of possible recoil energies of solar neutrinos is compared to the total energy range of the detector, the larger the gain in sensitivity from directional information. On the other hand, for the same number of background events Xenon can probe smaller cross sections.

It is not clear how a Xenon detector might be made directional. However, an additional motivation to pursue directionality is that ultimately very large Xenon detectors would be limited by the background of solar neutrino-electron elastic scattering events.  It is important to note that we do not take this background source into account in our simulations, although it is expected to become significant at the $10^{-48}$ cm$^2$ level~\cite{Baudis:2013qla}.  Additionally, the electron discrimination in Xenon detectors is less efficient than in other detectors, for example liquid argon. Directionality could aid in that discrimination considerably, because these events have a direction that points back to the Sun, allowing deeper cross sections to be probed.  Of course it is important to note that such large detectors would cost a great deal of money since Xenon is $\sim$\$1000 per kg at today's prices and only 25 tons are obtained from the air annually.

In this section we have presented results from detector simulations for which we fixed the number of expected neutrino events to 500 in order to estimate possible sensitivity limits in the presence of neutrino backgrounds. It was shown that cross sections beyond the discovery limit can be probed when directional information is taken into account. Directional detectors have significantly larger sensitivities for light dark matter masses. For a light target material this is also true for heavy dark matter. We will now move on and discuss how these limits behave as a function of exposure.

\subsection{\label{sec:max_sensitivity} Projected Sensitivity}

For both detector configurations that were presented in section~\ref{sec:results}, we choose three dark matter masses and for different exposures find the minimal cross section that can be tested at a 3$\sigma$ level with and without directional information. In this way we can find the exposure necessary to go beyond the discovery limit. As the dark matter masses we choose 6~GeV in order to see how directionality helps when the energy spectrum of solar $^8B$ neutrinos and dark matter are identical, 1000~GeV as a heavy dark matter mass and 30~GeV as a mass for which nondirectional experiments have close to maximal sensitivity. We expect the projected nondirectional sensitivity limits for the Xenon detector to flatten out below the limits from reference~\cite{Billard2013} by a factor of 2 because we assume half the neutrino flux uncertainties. For each dark matter mass and cross section we simulate $5 \times 10^3$ pseudoexperiments.

\begin{figure}
\includegraphics[height=7.cm]{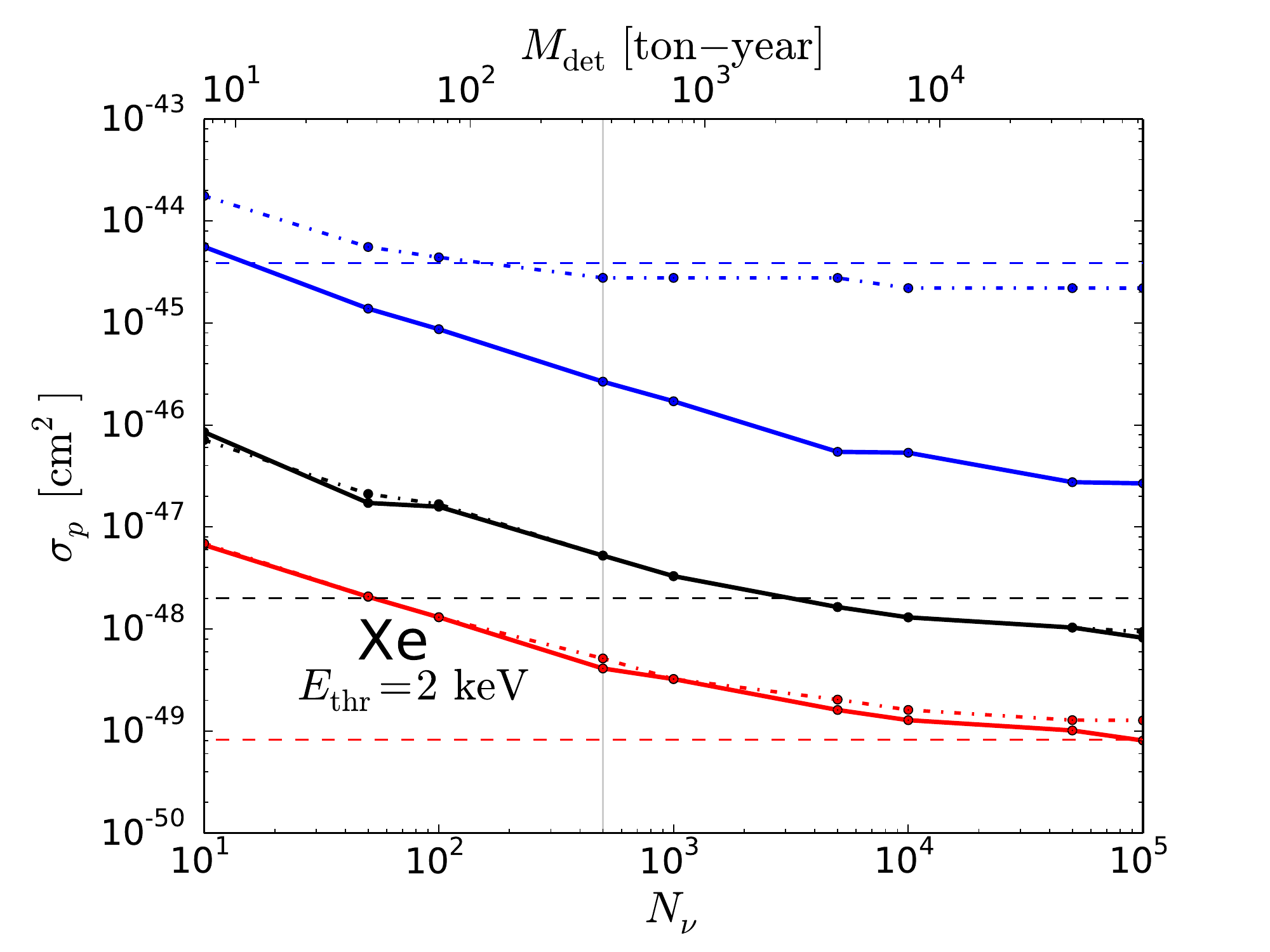}
\caption{\label{fig:sens_xenon}  Estimated sensitivity limits for a directional Xenon detector with a 2 keV energy threshold for a 6 GeV (blue), 30 GeV (red) and 1000 GeV (black) dark matter particle. The solid lines show directional detectors, the dash-dotted lines show nondirectional detectors. The horizontal dashed lines indicate the discovery limit of~\cite{Billard2013} for each dark matter mass. The vertical grey line shows the simulated detector of section~\ref{sec:results}}
\end{figure}

\begin{figure}
\includegraphics[height=7.cm]{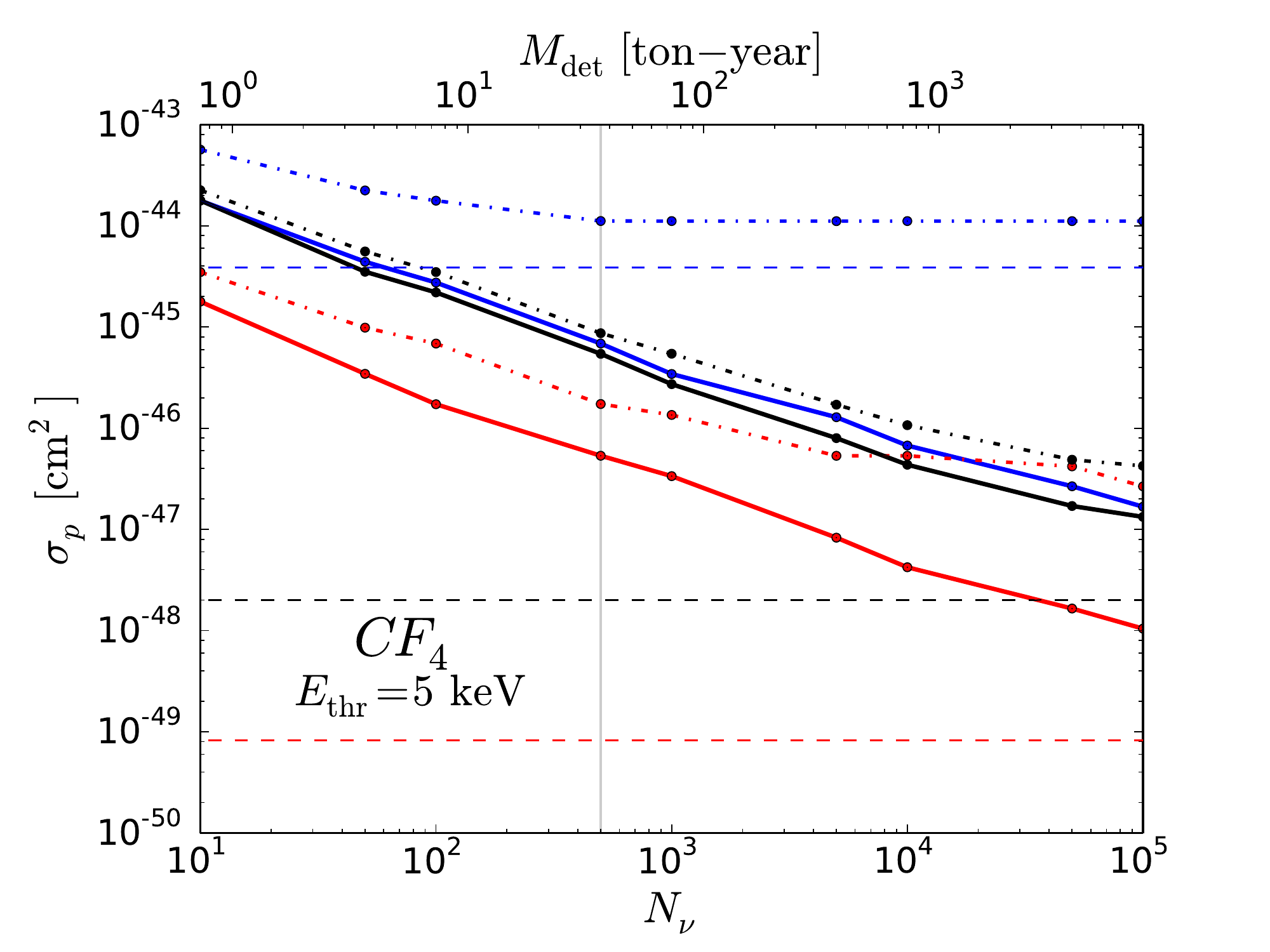}
\caption{\label{fig:sens_cf4} Estimated sensitivity limits for a directional CF$_4$ detector with a 5 keV energy threshold for a 6 GeV (blue), 30 GeV (red) and 1000 GeV (black) dark matter particle. The solid lines show directional detectors, the dash-dotted lines show nondirectional detectors. The horizontal dashed lines indicate the discovery limit of~\cite{Billard2013} for each dark matter mass. The vertical grey line shows the simulated detector of section~\ref{sec:results}.}
\end{figure}

We present the results for the Xenon detector in figure~\ref{fig:sens_xenon} and for the CF$_4$ detector in figure~\ref{fig:sens_cf4}. The projected sensitivity limits for directional detectors are presented as solid lines, nondirectional detectors are shown as dashed-dotted lines. The discovery limits of~\cite{Billard2013} are indicated as horizontal dashed lines. We color code the three different masses in blue (6~GeV), red (30~GeV) and black (1000~GeV).

As expected, for $m_{\rm DM} = 6$~GeV the sensitivity of nondirectional Xenon detector flattens out just below the discovery limit. This shows our agreement with earlier work~\cite{Billard2013} and shows directly the impact of improved knowledge on neutrino fluxes by a factor of two. For the nondirectional CF$_4$ detector (figure~\ref{fig:sens_cf4}) we see that the curve becomes flat already above the discovery limit. This is simply because the discovery limits were calculated for Xenon and we consider a different target material here. The solid blue line indicates possible limits if directional detectors were constructed. It is visible that the solar neutrino floor disappears for both target materials once there is a clear way to distinguish signal from background using directionality.

For the larger dark matter masses the directional and nondirectional limits are basically the same when Xenon is used as a target material, as we discussed in section~\ref{sec:results}. We see that the projected sensitivities reach a boundary close to the discovery limit of~\cite{Billard2013}. For the 1000 GeV WIMP this is slightly below the boundary as we expect from the reduced flux uncertainties. The 30 GeV sensitivity line stays just above the neutrino bound. We note our that the discovery limits in~\cite{Billard2013} were obtained with a 4~keV energy threshold, compared to a 2~keV threshold here and the different statistical approaches that were taken. Sensitivities for dark matter masses in this region show a dependence on the threshold energy. Besides, this change in threshold energy affects the neutrino energy spectrum significantly, which influences sensitivities of medium dark matter more than of heavy dark matter.

Because the directional and nondirectional projected sensitivities are close to identical, going beyond discovery limits for these masses with a heavy target material is only possible if the uncertainties of the neutrino fluxes are reduced and extremely large exposures become possible. Note here that our analysis does not include the $pp$ solar neutrino-electron scattering background, which is relevant for cross sections below $10^{-48}$ cm$^2$.

In figure~\ref{fig:sens_cf4} we show the sensitivity limits for a CF$_4$ detector. Similar trends are visible. Directional information allows to go beyond the discovery limits for light WIMPs even with imperfect flux knowledge. Directionality contributes significantly for the complete dark matter mass range and can help more than an order of magnitude in the cross section. For the light target material, we see a significant contribution from directionality even to the sensitivity of heavy WIMPs, but their discovery limits cannot be reached with the exposures we consider.

In the presence of backgrounds, the sensitivity is not expected to scale linearly with exposure. In a Poisson dominated regime, a scaling behavior as the square root of the number of background events is expected. Since we are using additional information here, coming from the energy spectrum, which might be identical for background and signal, and the directionality of the events, a better scaling behavior is possible. For the two detectors we simulated, the isotropic background of nonsolar neutrinos is smaller for the CF$_4$ compared to the Xenon case, resulting in a better scaling behavior. This, however, depends on the energy threshold.

In figures~\ref{fig:sens_xenon} and~\ref{fig:sens_cf4} the improvement in sensitivity is visible when directional information is included in addition to event time and recoil energy. In our analysis we find that time information adds only little on top of the sensitivity of a pure spectral analysis. For a 6~GeV dark matter particle in a CF$_4$ detector, e.g., we can find an improving effect  of more than 10\% on the sensitivity when measuring annual modulation only if there are about $10^3$ background events. We see that annual modulation becomes important only for large background rates, and that the impact of directional information is much larger.

The estimated sensitivities of directional detectors also depend on the chosen angular and energy resolutions. We find the angular resolution to be the more important one. For a 6~GeV dark matter particle decreasing the energy resolution by a factor 2 would leave the sensitivity unchanged up to a few percent. The same change in the angular resolution, however, would reduce the sensitivity by a factor $\sim 3$ for 500 background events or even a factor $\sim 5$ if there are 5000 background events.

These plots show that in principle there is no solar discovery limit for direct dark matter searches if directional detectors are constructed. Going beyond the discovery limit for a 6~GeV dark matter particle is possible for an exposure of approximately 5 ton-years for a directional CF$_4$ and around 10 ton-years for a directional Xenon detector. For this dark matter mass, a directional experiment can reach the discovery limit with an exposure which is smaller by about an order of magnitude compared to the nondirectional case. Directionality has more impact the lighter the dark matter particle. For events to be above the energy threshold, the incoming light dark matter particles need a large velocity and, hence, have a clear arrival direction from Cygnus A. For a light target material directionality also adds to the sensitivity of heavy dark matter candidates.

\section{\label{sec:end} Discussion and Conclusions}
In this work we looked at future sensitivities of direct dark matter searches when irreducible neutrino backgrounds from coherent neutrino-nucleus elastic scattering is present. We investigated how time, recoil energy and directional information can help one distinguish signal from background. To do so, we performed a hypotheses test, as explained in section~\ref{sec:stats}, and demanded separations of the two hypotheses at 90\% confidence and three sigma level. 

For the simulated detectors we assumed moderately optimistic energy thresholds and energy efficiency behavior as well as realistic smearing in the energy and angular resolution. In order to see how the target mass influences the searches, we looked at tetrafluoromethane, CF$_4$, as a light and Xenon as a heavy target material. For CF$_4$ there are detector technologies that measure the recoil track of the nucleus, whereas they have not yet been developed for Xenon. 

In section~\ref{sec:detectors} we presented two dimensional probability distributions in recoil energy and event angle, $\theta_{\rm sun}$, for neutrino and dark matter events. In figure~\ref{fig:dm_pdf} we showed the distribution for light dark matter and pointed out that it peaks at large values of $\cos \theta_{\rm sun}$. We discussed how the position of this peak evolves over the year due to the motion of the Earth around the Sun. We remarked that the lightest dark matter particles that a detector is sensitive to, need to have a large incoming velocity such that their arrival direction points back to Cygnus A.

In the same section we presented in figure~\ref{fig:neutrino_pdf} the corresponding distribution for the neutrino events. The way we defined the event angle removes any time dependence of this distribution. Compared to dark matter, we noted that the solar neutrinos peak at small recoil energies and at $\cos \theta_{\rm sun} \approx -1$. The nonsolar neutrinos were assumed to have an isotropic distribution in the detector frame and act as a smooth background for the distribution.

In section~\ref{sec:results} we simulated one detector for each target material and fixed the exposure such that there are 500 expected neutrino events. We found that with directional information cross sections beyond the neutrino discovery limit may be probed. For the light target material we see that directional information is helpful for the complete dark matter mass range, whereas for the heavy target nuclei, directional and non-directonal detectors will give the same limits for heavy dark matter. In both cases, directional detectors can test more than an order of magnitude smaller cross sections compared to nondirectional detectors for some light dark matter masses.

We projected possible sensitivities as a function of exposure in section~\ref{sec:max_sensitivity}. In figure~\ref{fig:sens_xenon} and~\ref{fig:sens_cf4} we saw that directional information removes the solar neutrino discovery limits and is especially useful for light dark matter. For a 6~GeV dark matter particle an exposure of approximately 5 ton-years for a directional CF$_4$ and around 10 ton-years for a directional Xenon detector is sufficient to go beyond the discovery limit of nondirectional detectors. The limit is reached with an exposure  that is about an order of magnitude smaller compared to the nondirectional case. If a light target material is used, a gain in sensitivity exists also for heavy dark matter candidates, but very large exposures are needed to reach their discovery limits. We noted that perfect flux knowledge would also remove any discovery limit and conclude that there is no neutrino bound for directional dark matter searches.

\begin{acknowledgments}
This work was supported by the ERC and the STFC.  We are grateful to Peter Fisher for collaboration in the early stages of this project and commenting on the manuscript. Also, we wish to thank Tevong You and Thomas Richardson for helpful discussions, Julien Billard for valuable comments on the draft and Nigel Arnot for help with software, hardware and network problems.  MF is grateful for the hospitality of NORDITA during the "What is dark matter'' program where many relevant interesting conversations took place.
\end{acknowledgments}
\bibliography{nulimit}
\end{document}